# MIT Lincoln Laboratory: A Case Study on Improving Software Support for Research Projects

Daniel Strassler
MIT Lincoln Laboratory
Lexington, MA, USA
daniel.strassler@ll.mit.edu

Gabe R. Elkin
MIT Lincoln Labortatory
Lexington, MA, USA
gabee@ll.mit.edu

Curran Schiefelbein
MIT Lincoln Labortatory
Lexinton, MA, USA
curran.schiefelbein@ll.mit.edu

Daniel Herring
MIT Lincoln Laboratory
Lexington, MA, USA
dherring@ll.mit.edu

Ian Jessen
MIT Lincoln Laboratory
Lexington, MA, USA
ian.jessen@ll.mit.edu

David Johnson
MIT Lincoln Labortatory
Lexington, MA, USA
davidj@ll.mit.edu

Santiago A. Paredes
MIT Lincoln Labortatory
Lexinton, MA, USA
santiago.paredes@ll.mit.edu

Tod Shannon
MIT Lincoln Laboratory
Lexington, MA, USA
tod@ll.mit.edu

Jim Flavin
MIT Lincoln Laboratory
Lexington, MA, USA
jflavin@ll.mit.edu

*Abstract*— Software plays an ever-increasing role in complex system development and prototyping, and in recent years, MIT Lincoln Laboratory has sought to improve both the effectiveness and culture surrounding software engineering in execution of its mission. The Homeland Protection and Air Traffic Control Division conducted an internal study to examine challenges to effective and efficient research software development, and to identify ways to strengthen both the culture and execution for greater impact on our mission. Key findings of this study fell into three main categories: 1 - project attributes that influence how software development activities must be conducted and managed, 2 - potential efficiencies from centralization, 3 – opportunities to improve staffing and culture with respect to software practitioners. The study delivered actionable recommendations, including centralizing and standardizing software support tooling, developing a common database to help match the right software talent and needs to projects, and creating a software stakeholder panel to assist with continued improvement.

*Keywords— software, software developers, software engineers, DevOps, DevSecOps, Research Software Support, case study*

## I. INTRODUCTION

MIT Lincoln Laboratory (MIT LL) has a long history of developing novel technical solutions in support of national security. While software development has always been present, it was often a significantly lower portion of project execution, and the nature and complexity of software development activities have evolved alongside the growth of the software industry and open source communities. Over the past two decades, software development has seen a significant increase in Laboratory project work and deliverables. Accordingly, the percentage of staff at MIT LL with computer science / engineering degrees has doubled since 2000 (Figure 1). Recognizing this trend, the Laboratory has taken a keen interest in enhancing the effectiveness of software engineering practices and culture, with the goal of supporting and bolstering this core competency.

In addition, as responsible stewards of US research and development funds, it is incumbent on staff to provide efficient and impactful effort at the ever-increasing pace national security requires for all projects, software development projects included. To that end, the Laboratory has performed several internal studies to identify opportunities to increase both the impact and speed of the Laboratory's software development.

This paper will discuss a software study conducted within the Homeland Protection and Air Traffic Control division. The study had four objectives:

- to understand and assess the software development communities and practices within the division
- to discover and document the challenges experienced across these communities

- to evaluate alternative models for the organization, management, and capabilities needed to support these communities
- to recommend an organizational approach that supports the division software community needs and addresses the challenges.

The study was conducted in three phases: data collection (internal and external), analysis of the data collected, and development of a final report with recommendations. It was led by two managers from different groups within the division, with oversight provided by a division level principal staff member. The team included a technical staff member from each group within the division to represent their group's perspectives and support the study's execution. All members, leadership and technical, have decades of experience in the practice and management of software engineering.

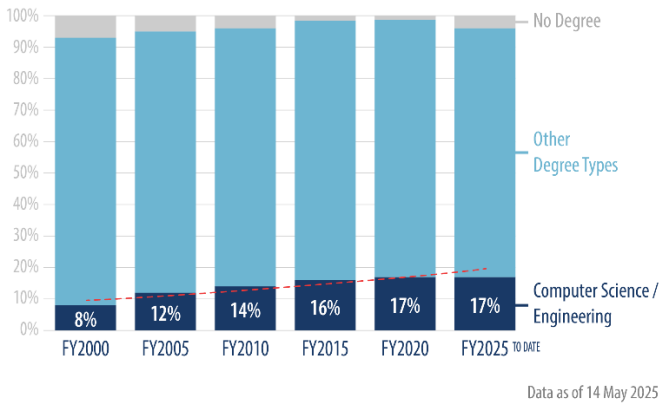

Fig. 1. Percentage of MIT Lincoln Laboratory technical staff with Computer Science or Computer Engineering degrees, 2000-2025.

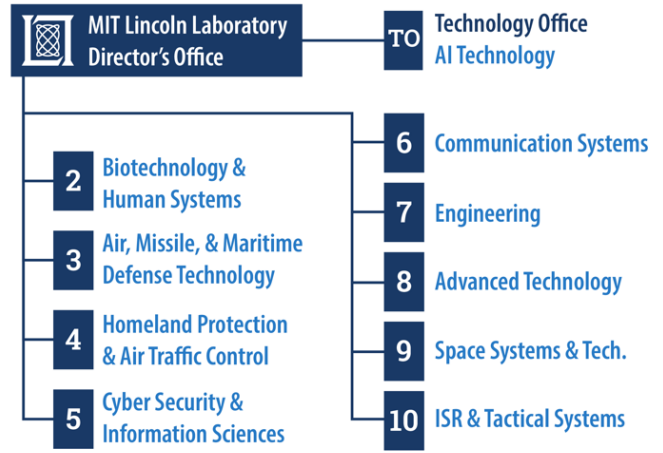

Fig. 2. MIT Lincoln Laboratory Technical Divisions

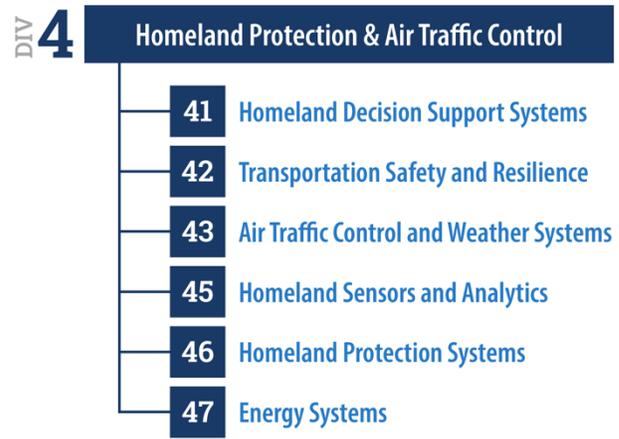

Fig. 3. Division 4, Homeland Protection & Air Traffic Control Groups

II. ORGANIZATIONAL AND OPERATIONAL STRUCTURE OF THE ENTERPRISE

To understand the scope of this study, an understanding of MIT Lincoln Laboratory's organizational and operational structure is necessary. The Laboratory is organized into technical divisions and operational departments under the governance of the Director's Office. Each of the technical divisions has a mission focus, as shown in Figure 2. Groups within a division are likewise differentiated by their technology- or mission-centric focus areas.

The Homeland Protection and Air Traffic Control Division, which conducted this study, is comprised of six groups ranging in size from roughly 25-75 employees, and with focus areas outlined in Figure 3. The groups are responsible for executing portfolios of projects, which range in quantity, size, and duration, per sponsor needs. An individual project's level of effort may range from half a full-time equivalent (0.5 FTE) three-month study up to dozens of FTEs over multiple years.

From an operational perspective, the Laboratory's work on technical and mission challenges has intrinsic characteristics which influence how projects are managed and executed. The sponsors of the work impose diverse requirements, such as deployment environments or programming languages used. For example, one project may have the requirement to deploy to a network connected data center where there are few resource constraints, while another project may have the requirement to deploy to a low Size, Weight, and Power (SWaP) device in a rugged or remote environment. The tooling, technologies, and execution of these projects would likely have minimal overlap due to their differing requirements. The projects the Laboratory executes also are part of a larger ecosystem of Commercial Off-the-Shelf (COTS), Government Off-the-Shelf (GOTS), and custom software and hardware, with which interoperability or direct integration is desirable. This business environment presents challenges that may result in a different decision space than that of an organization or company with full decision power over its software product and deployment environment.

## III. Internal Data Collection

There were three objectives of the data collection phase. First and foremost, the study sought to gather metrics at the group level to understand the scopes and attributes of current projects involving software development, tooling in use for software development, and culture surrounding software development. This was accomplished through a survey filled out by each group's technical representative, informed by their group's staff. One set of questions focused on the projects and staff supporting the execution of the group's software projects. From these questions, the study identified two key metrics: the number of projects within an individual group categorized by the number of staff contributing to them, and the number of staff contributing to software development within the group. A second set of questions focused on the work products ("deliverables"), deployment environments, and technologies associated with packaging and deploying the interim and final artifacts. The final set of questions gathered information about the development process, including project management, version control, testing, and DevSecOps (Development, Security, Operations) processes and tooling.

The second objective was to examine how a representative subset of individual projects managed their software engineering activities, e.g., version control and build automation tooling, project management processes like Sprint or SCRUM, and their perceived effect on software development. The study also explored how sponsor requirements, funding, staffing, deliverables, or other considerations affected those decisions and processes. To accomplish this, the study team conducted interviews with software practitioners for a subset of projects, with a common set of questions to guide the discussion and allow for freeform responses. The questions focused on execution challenges and pain points with software development tooling and staffing, e.g., aligning project timelines and staff availability, or balancing competing priorities for access to specialized hardware. Each of these interviews involved the project lead and often multiple project team members.

Finally, the internal data collection effort sought to catalog existing capabilities and technologies developed in each group that support software practitioners and their projects in response to the challenges they described in the team discussions. This analysis phase identified a gap in cataloging the specific tools which provide capabilities like static code analysis and artifact repository services. To resolve this gap, a follow-on data collection was performed to acquire a listing of vendor tooling used to satisfy a key objective of identifying recommendations for cost saving and efficiency.

## IV. External Document Review

The study also reviewed external drivers for change in software practices, including trends in government requirements for software development, deployment technology transfer, and industry best practices. This review focused on items referenced from the internal data collection to scope the effort appropriately. Some documents provide guidance but not implementation details regarding software development to the Department of Defense and its contractors, e.g., the DoD Enterprise DevSecOps Strategy Guide [1]. Other documents, e.g., NIST-provided SP-800-XXX [2] standards, provide guidance on implementation of security controls. A limitation here was that some government-specific documentation was difficult to include due to dissemination limitations.

## V. Analysis

The analysis phase involved aggregating and reviewing both the survey and interview results. This information was then combined with the external document review to identify common themes that could be used for actionable recommendations. Three categories emerged during the review of the material. The first one, project attributes, are characteristics of software projects that influence the management and execution at the level of writing code, collaborating internally and externally, and producing and deploying/distributing artifacts such as executables and documentation. The second category, centralization, identifies which tools, processes, and technologies are commonly used (whether centrally managed or duplicated across the enterprise). The third category, personnel and culture, includes findings relating to staff skills and project staffing, specifically in regard to staff with software experience.

### A. Findings: Project Attributes

There were two key findings under the project attributes category. The first was that software development within the division is more diverse than it is similar, due to the deployment environment, technology, or integration requirements not within the projects' control. Some of the variability is depicted in Figure 4, which shows a subset of the survey responses and interview results for objectives, deployment environments, languages used, and chip architecture. The projects have a range of objectives from internal use software for ad-hoc tasking to operational systems. While this finding is not directly actionable, it does predict that trying to achieve greater efficiencies through standardization on a limited set of technologies would have a restrictive effect on the teams' ability to deliver on requirements.

| Objectives | Deployments | Languages | Architecture |
|---|---|---|---|
| Internal Use | Embedded | C/C++ | X86 |
| Proof-of-Concept | Desktop | Python | GPU |
| Functional Prototype | Server | Java | ARM |
| Operational Prototype | Cloud | Javascript | Other |
| Operational System | Mobile Device | Matlab | |
| | LLSC | Rust | |
| | | Tcl/Tk | |

Fig. 4. Survey and interview results for project details

The second finding was that team size at the project level, and the composition of these project teams, vary within each group, as shown in Figure 5. Team size was determined by the number of staff contributing to software products. For a staff member to be counted, they only needed to be contributing and were not required to be working on the project full-time. The team sizes were bucketed into 1-2 staff, 3-5 staff, and 6 or more staff. The wide variability in team size and composition predicts that a top-down approach to process standardization would have mixed benefit due to different project team sizes and communication patterns within teams.

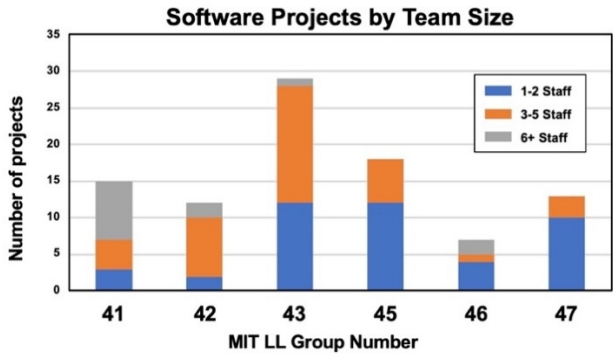

Fig. 5. Software projects team size across groups in Division 4

*B. Findings: Centralization*

There were three key findings under the centralization category. The first is that MIT Lincoln Laboratory provides relatively limited enterprise support for software, including development tools such as an internal GitHub Enterprise service and centrally managed licenses for popular integrated development environment (IDE) software. At the time of the study, additional DevSecOps capabilities or services were not provided at the enterprise level or were only available in a limited "pilot" capability. Therefore, software process decisions and tooling to support those decisions were relegated to the group or project level. On small teams, the decisions have often been delegated to individual developers, who out of necessity establish conventions, and purchase DevSecOps solutions, local to their team.

The second finding under centralization is a consequence of the first. Each group, and in some cases projects, have organically developed and deployed tools to support their software processes. These organic capabilities have enabled teams to meet their project requirements. It was also discovered that knowledge sharing was occurring in small communities across projects and groups with some efficiencies realized through shared tooling, but with limited discoverability across the division or enterprise.

The third finding was that many projects and groups have similar requirements for the software developed and deployed, including documentation, security measures, and infrastructure. In several cases, common tooling was used to satisfy the requirements; however, each group was running their own instance. For example, three groups were all using the same static analysis tool, SonarQube, but had their own instance they were maintaining. In other cases, the requirement or functionality was the same but the tooling used was different. For example, some groups were using Jenkins to execute their Continuous Integration (CI) pipelines while others were using TravisCI.

*C. Findings: Personnel and Culture*

There were two key findings under the personnel and culture category. The first regarded staffing, i.e., the timely availability of staff with a needed skill set for a project. During several interviews, team leads mentioned having no easy way to discover whether someone with a specific skill set was available. The current method of discovering staff with needed skill sets was word of mouth using pre-existing social networks. For example, a team lead looking for a staff member with strong PyTorch experience would reach out to staff they had worked with in the past to see whether anyone knew another staff member with that experience. There was no known search tool or central repository mentioned during the interviews. Also, recruitment and retention presented as a compounding challenge. Specific skill sets were mentioned in multiple interviews as being in short supply but high demand, e.g., Kubernetes experience. During the interviews, the difficulty of competing with companies like Meta, Alphabet, and Amazon were mentioned several times with respect to recruiting and retaining software engineers.

The second finding was that discovering a community of experts on a specific software topic was difficult. Similar to the first finding, the interviewees discussed difficulty in finding staff with specific topic knowledge to consult. As above, the lack of a known search tool or central repository of staff with specific software knowledge was a challenge partially met by using pre-existing social networks.

VI. FINAL REPORT RECOMMENDATIONS

The final phase of the study was to provide a report and out-brief with recommendations for the division to consider. After the out-brief, the study team and division leadership reviewed the full report recommendations with an eye toward near term actions within the division as well as opportunities for broader, Laboratory-wide engagement.

Recommendations were categorized as in prior sections, with the exception of project attributes. The project attributes category held no actionable recommendations due to the findings highlighting factors beyond the Laboratory's control or desire to control.

*A. Centralization Recommendations*

- Establish a division software stakeholder panel with representatives from each group and division leadership, staffed initially by the division software study participants.

- Empower the division software stakeholder panel members with responsibility to solicit and represent their respective groups software interests / needs for division level decisions and promote development and adoption of common software support capabilities within their groups.

- Expand division-level DevSecOps capabilities to enable projects with common services and solutions through the consolidation of duplicate services and similar capabilities.

- Share solutions and automation though infrastructure as code (IaC) for common requirements, in order to create a centralized solution and knowledge repository.

- Engage with the Knowledge Services enterprise team, an internal organization responsible for research and operational support, to increase discoverability and collaboration regarding existing software, models, data, and documentation. Contribute to standardization efforts regarding software knowledge management and discoverability.

*B. Personnel and Culture Recommendations*

- Create a searchable software skills database in collaboration with Human Resources, in order to facilitate matching project needs to staff skills, and request that staff update the database on a biannual basis.
- Explore the development of career enhancement opportunities for software engineers.
- Develop a software focused speaker series with external speakers from US government, industry, and academia.
- Engage with Human Resources to improve recruitment and retention of software professionals.

## VII. POST REPORT PROGRESS

After the final report and recommendations were provided to the division office, there have been multiple efforts at both the division level and the broader Laboratory level toward improving support for software development. To maintain alignment with the prior sections, the progress since the report will be categorized similarly to the recommendations section.

*A. Centralization*

- An enterprise-wide DevSecOps Collaborative community of interest was founded with meetings every other week, attended by the newly formed enterprise DevSecOps team and stakeholders from multiple divisions, including managers and engineers.
- The DevSecOps Collaborative developed an enterprise roadmap for enterprise-level tooling and capability, incorporating stakeholder feedback.
- An enterprise GitLab instance was deployed, consolidating multiple division and group level instances. This reduced support overhead and increased discoverability.
- An enterprise JFrog Artifactory instance with X-Ray capability was deployed, consolidating multiple division and group level artifact repository services for software and related products, such as JARs, Python Wheels, Docker images, and Software Build of Materials (SBOMs).
- An enterprise SonarQube instance was deployed, consolidating multiple instances of division and group level instances used for static code analysis and reporting to meet sponsor requirements for deployment approvals.
- An enterprise OpenText Fortify instance was deployed, consolidating multiple group level instances and decreasing licensing costs.
- Efforts are underway to develop a policy codifying naming conventions and implementation templates, in order to increase discoverability and employment of reusable code.
- Shared Infrastructure as Code (IaC) repositories were created for automated deployment of development resources and services.
- An initial implementation of templates for Continuous Integration / Continuous Deployment (CI / CD) pipelines was shared for common languages and deployment targets using enterprise services.

*B. Personnel and Culture*

- A Laboratory study was initiated to identify internal resources and commercially available products for talent management. The intent is to include functionality akin to a skills database, community of interest formation, and professional profiles similar to LinkedIn.
- Communication was reviewed at the Division level, regarding existing education benefits and existing internal training resources, to encourage and remind staff of these opportunities.

ACKNOWLEDGMENT

The authors would like to acknowledge and thank our division heads, Dr. Jennifer Watson, Dr. Christopher A. D. Roeser, Dr. Jonathan Pitts, and Dr. James K. Kuchar for sponsoring this study, being supportive and receptive to the recommendations, and helping us to share it with a broader audience outside the Laboratory so others could learn from it.

APPENDIX

*C. Survey Questions*

1. How many software development projects/programs do you have within the group?
2. How many programs in the group do you have that are 1 or 2 software contributors (individuals not FTEs)
3. How many programs in the group do you have that are 3 to 5 software contributors (individuals not FTEs)?
4. How many programs in the group do you have that are 5+ software contributors (individuals not FTEs)?
5. Approximately how many individuals within the group are contributing to software development?
6. Is software development growing or declining within your group over the past few years?
7. Do you anticipate this trend to continue into the next year?
8. What are your common software project deliverables?
9. What software development process(es) do you use?
10. What source code version control tools do you use within the group? (git, SVN, etc.)
11. What artifact repository/stores do you use within the group? (Artifactory, Nexus, etc.)
12. What issue/task tracking tools do you use within the group? (JIRA, Github issue, etc.)

13. What collaboration/communication tools do you use within the group? (Confluence, wikis, Sharepoint/Teams, chat, etc.)
14. What scanning tools do you use within the group? (Static/Dynamic code analysis, container, fuzzers, etc.)
15. What testing processes and tooling do you perform within the group? (Unit, Integration, regression, Automation vs Manual, etc.)
16. What build and automation tooling do you use within the group? (Maven, Jenkins, TeamCity, etc.)
17. What is the target deployment platform and environment? (embedded, desktop, server, cloud, mobile, etc.) If you utilize commercial clouds, can you list which? (AWS, AZURE, GCP, etc.)
18. If you utilize commercial clouds, can you list which? (AWS, AZURE, GCP, etc.)
19. What software packaging does your group use? (JARs, Exes, Wheels, etc.)
20. What orchestration technologies does your group use? (docker, kubernetes, ansible, etc.)
21. What chipset/architectures do you utilize within the group?
22. What code branching and code/design review process do you follow within the group?
23. What requirements management tools do you use within the group? (DOORs, Jama, etc.)
24. What are your DevSecOps processes?
25. What languages, frameworks, libraries, or other notable technologies / stacks are critical to your software development?
26. What topics related to software development are you most interested to see addressed?

*D. Interview Questions*

1. What are the software development challenges you have?
2. What specific software challenges do you want help with?
3. How do you use tooling for software development?
4. What supporting roles or skillset do you require for your software development programs?
5. What are your pain points and how do you get around them?
6. How are your software developers acquiring their knowledge? Is there support you want with this?
7. What staffing challenges do you have regarding software development?
8. What else do you want to tell us about software development?